# Coherent coupling of dark and bright excitons with vibrational strain

Ryuichi Ohta[1], Hajime Okamoto[1], Takehiko Tawara[1], Hideki Gotoh[1], and Hiroshi Yamaguchi[1]

[1]NTT Basic Research Laboratories, NTT Corporation,

3-1 Morinosato Wakamiya, Atsugi-shi, Kanagawa 243-0198, Japan

In many physical systems, there are specific electronic states called "dark states" that are protected from the rapid radiative decay imposed by the system symmetry [1-4]. Although their long-lived nature indicates their potential for quantum information[1,2,5,6] and spintronic[7] applications, their high stability comes at the expense of optical accessibility. Breaking the symmetry by using magnetic[8,9] and electric fields[7,10] has been employed to hybridize dark and bright states thus making them optically active, but high-frequency and on-chip operation remains to be developed. Here we demonstrate the strain-induced coherent coupling of dark and bright exciton states in a GaAs mechanical resonator. The in-plane uniaxial strain breaks the rotational symmetry of the crystal, allowing the dark states to be optically accessible without any external fields. Such dark-bright coupling is tailored by the local strain distribution, which enables the coherent spin operation[11,12] in the gigahertz regime and opens the way to on-chip excitonic quantum memories and circuits.

Optical transition in two-level systems is generally governed by the conservation law of total angular momentum. To optically excite (relax) two-level systems, the total angular momentum must be conserved between the initial and final states. In the particular case of solid state systems, the optical transition obeys the spin polarizations of electrons and holes. For instance, electrons in the conduction band and holes in the heavy-hole band with antiparallel spins are optically active, and are called bright excitons, and those with parallel spins are optically inactive, and are called dark excitons. Although dark excitons are of great interest owing to their long lifetimes[5-7,13], direct optical access is technically challenging[14,15]. A magnetic field is commonly used to break the rotational symmetry provoking the coherent coupling of dark and bright excitons[8,9]. Such hybridization allows the dark excitons to become optically active. However, a large magnetic field is only induced statically and uniformly, which limits the dynamic and local control of the dark excitons.

A strain field breaks the rotational symmetries of the crystals, and causes intermixing of dark and bright excitons via deformation potentials. In contrast to a magnetic field, a strain field is dynamically and locally controlled with piezoelectric materials[16,17]. The strain effects for exciton states are described by the Pikus-Bir Hamiltonian[18,19] as follows.

$$\begin{pmatrix} P_\epsilon + Q_\epsilon & L_\epsilon & M_\epsilon & 0 \\ L_\epsilon^* & P_\epsilon - Q_\epsilon & 0 & M_\epsilon \\ M_\epsilon^* & 0 & P_\epsilon - Q_\epsilon - \varepsilon_{ex} & -L_\epsilon \\ 0 & M_\epsilon^* & -L^* & P_\epsilon + Q_\epsilon - \varepsilon_{ex} \end{pmatrix} \begin{pmatrix} |BX_{HH}\rangle \\ |BX_{LH}\rangle \\ |DX_{LH}\rangle \\ |DX_{HH}\rangle \end{pmatrix} \quad (1)$$

$$P_\epsilon = (a_c - a_v)(\epsilon_{xx} + \epsilon_{yy} + \epsilon_{zz}) \quad (2a)$$
$$Q_\epsilon = -b(\epsilon_{xx} + \epsilon_{yy} - 2\epsilon_{zz})/2 \quad (2b)$$
$$L_\epsilon = d(\epsilon_{xz} - i\epsilon_{yz}) \quad (2c)$$
$$M_\epsilon = \sqrt{3}b(\epsilon_{xx} - \epsilon_{yy})/2 \quad (2d)$$

Here, BX (DX) stands for bright (dark) excitons, where the subscript of HH (LH) specifies a heavy-hole (light-hole) band. $P_\epsilon$, $Q_\epsilon$, $L_\epsilon$, and $M_\epsilon$ are strain induced perturbations, which are derived from strain tensors ($\epsilon_{ij}$) and the deformation potentials of GaAs ($a_c$, $a_v$, $b$, $d$). $\varepsilon_{ex}$ is the exchange energy of an electron and a hole. Uniaxial in-plane strain ($\epsilon_{xx} \neq \epsilon_{yy}$) provides non-zero off-diagonal components $M_\epsilon$, which cause the inter-band mixing of BX and DX, whereas $M_\epsilon$ becomes negligible with isotropic in-plane strain ($\epsilon_{xx} = \epsilon_{yy}$), which can be caused by the lattice mismatches in heterojunctions. Therefore, uniaxial strain breaks the four-fold rotational symmetry of the zincblende structure, allowing the hybridization of different total angular momentum states, i.e. bright excitons and dark excitons as shown in Fig. 1(a).

To generate time-varying uniaxial in-plane strain, we adopted a micro mechanical resonator based on GaAs, which greatly enhances the strain amplitude because of its quality factor. Figure 1(b) show scanning electron microscope images of this resonator. The sample preparation is detailed in Method. Figure 1(c) shows the calculated energies of the bright and dark excitons in the HH and LH bands with various mechanical displacements, where the strain tensors are calculated with the finite element method

(FEM) (see SI). We take account of the contribution from the intrinsic isotropic strain induced by the epitaxial mismatch, which splits the exciton energies of the HH and LH bands when the cantilever is in the equilibrium position[20]. The applied uniaxial strain causes the additional energy splits and the mixing of the bright excitons in the LH band and the dark excitons in the HH band when the mechanical displacement exceeds 20 nm.

Figure 2(a) shows the experimental setup used in this study, which consists of a Ti:Sa pump laser, a charge coupled device (CCD), and a Doppler interferometer. The exciton energies were characterized with a photoluminescence (PL) measurement, while the mechanical motion was simultaneously measured with the interferometer. The characterizations were performed at 7.2 K. The experimental setup is explained in detail in Method. Figure 2(b) shows a typical PL spectrum obtained at the mid-length of the resonator. The position dependence of the PL spectra is discussed in SI. The three higher (lower) energy peaks are attributed to donor (acceptor) bound excitons[21]. To induce time-varying strain at that position, the 2nd order flexural mode oscillating at 1.04 MHz was driven by a piezo actuator. Figure 2(c) shows the frequency response of this mechanical mode at various drive voltages, where the inset shows the strain distribution calculated by FEM. A mechanical $Q$ of 32,000 is extracted from the linewidth. The vibrational amplitude at the resonance frequency and the corresponding uniaxial strain are shown in Fig. 2(d). The mechanical nonlinearity gradually saturates the amplitude at an actuation voltage higher than 6 mV$_{pp}$, while the corresponding strain is high enough to cause coupling as estimated from Fig. 1(c).

We employed stroboscopic PL measurement[16,22,23] to investigate the effects of mechanically induced strain on the bound exciton energies. The pump laser was shaped into a periodic rectangular pulse with an acousto optical modulator (AOM), whose repetition rate was synchronized to the mechanical frequency. By changing the relative phase between the pump pulse and the mechanical motion, PL spectra under time-varying strain were obtained with a CCD. Figure 3(a) shows stroboscopic PL spectra of bound exciton states for a drive voltage of 12 mV$_{pp}$. Figure 3 (b)-(g) are cross-sectional PL spectra of acceptor/donor bound excitons at 0 (d/g), 320 (c/f), and 480 ns (b/e), respectively. Each spectrum is fitted by a Lorentz function with the peaks labeled I to VI, and their positions are traced as dashed lines in Fig. 3(a). The vibrational strain sinusoidally modulates peaks I and II, whereas peak III is less affected. These behaviors reproduce the energy shifts of acceptor bound excitons characterized by static strain experiments[24,25]. The vibrational strain also modulates the energy difference of peaks V and VI, which originate from the bright exciton states bound by donors. The most interesting feature can be seen around 320 and 640 ns, where peak V shows the avoided crossing with peak IV. Peak IV is attributed to dark excitons in accordance with ref 26. In fact, its PL intensity vanishes when the energy is detuned from peak V. The avoided crossing is a clear sign that the bright and dark excitons are coherently coupled by mechanically induced strain.

To investigate the mechanically hybridized exciton states quantitatively, we numerically calculated the energies and PL spectra of bound excitons with the Pikus-Bir Hamiltonian. We derived the values of $P_\epsilon, Q_\epsilon, L_\epsilon,$ and $M_\epsilon$ from the measured mechanical amplitudes and the calculated strain tensors with FEM. Figure 4(a)-(d) show the measured PL spectra and calculated dark and bright bound exciton energies (dashed lines) at drive voltages of 2, 6, 12, and 16 mV$_{pp}$. Figure 4(e)-(h) show the calculated PL spectra, where the linewidths of the dark and bright excitons are 70 μeV (see SI). The good agreement between the experiments and the theoretical analysis confirms the hybridization of the dark and bright bound excitons via mechanical oscillation. The coupling strength ($g_{DB}$) is estimated to be about 50 μeV, which is of the same order as the linewidths.

The $\boldsymbol{k} \cdot \boldsymbol{p}$ model analysis indicates that $g_{DB}$ can be further improved by tailoring $\epsilon_{xx}$ and $\epsilon_{yy}$. For instance, we can independently control $Q_\epsilon$ and $M_\epsilon$ in a drum-shaped mechanical resonator, and increase $g_{DB}$ to 1 meV, which corresponds to a cooperativity ($C = 4g_{DB}^2/\gamma_B\gamma_D$) of 800 thus allowing the coherent manipulation of the two states (see SI). Moreover, the Pikus-Bir Hamiltonian suggests that the off-diagonal terms of $L_\epsilon$, namely the shear strain of $\epsilon_{xz}$ and $\epsilon_{yz}$, cause additional coupling between $|BX_{HH}\rangle$ and $|BX_{LH}\rangle$ ($g_{BB}$), and $|DX_{LH}\rangle$ and $|DX_{HH}\rangle$ ($g_{DD}$). The shear strain is generated by torsional oscillation and is concentrated near the side edges of the resonator. We evaluated the additional exciton-exciton couplings in the 2nd order torsional mode of this resonator shown in Fig. 5(a). Figure 5(b) shows the calculated population of each bound exciton state, which describes the three avoided crossings representing the coherent couplings in multiple dark and bright excitons (see SI).

In conclusion, we demonstrate the coherent coupling of dark and bright bound excitons and modulate their optical accessibility using time-varying strain in a mechanical resonator. The origin of the dark-bright coupling is the breaking of the rotational symmetry of the system. The good agreement between the experimental results and the $\boldsymbol{k}\cdot\boldsymbol{p}$ theory calculation confirm that the vibrational strain causes the hybridization of the two states. Further improvement in the couplings ($g_{DB}$, $g_{BB}$, and $g_{DD}$) is possible by tailoring the strain distribution. Although the coherent coupling of dark and bright excitons has been demonstrated with a magnetic field, this mechanical approach has significant advantages as regards the modulation speed of the coupling strength. This device was operated at around 1 MHz, and this will be increased to several gigahertz with current MEMS technologies. In this regime, the coupling can be more rapidly switched than the decay of the bright excitons, which will provide the coherent control of the single spin in dark and bright excitons[11,12] on a solid-state platform. This mechanical approach can be widely applied to any solid-state two-level system, such as quantum dots[22,23,27,28], nitrogen-vacancy centers[29,30], and impurities in semiconductors, and makes it possible to realize on-chip excitonic memories and circuits.

## Method

The mechanical resonator consists of three layers of 100-nm-thick *n*-doped $Al_{0.3}Ga_{0.7}As$, undoped 400-nm-thick GaAs, and 5-period undoped $Al_{0.3}Ga_{0.7}As$/GaAs superlattices. They were grown by molecular-beam epitaxy on a semi-insulating GaAs substrate with a 3-μm-thick $Al_{0.65}Ga_{0.35}As$ sacrificial layer. The mechanical resonator was fabricated by using photolithography, wet etching with $H_3PO_4$ and $H_2O_2$, and under-etching with HF. The fabricated resonator was 37 μm long, 20 μm wide, and 600 nm thick. The resonator was attached to a piezo actuator, which drove it electrically at the resonance frequency, and placed in a vacuum chamber in a closed-cycle cryostat (attoDRY 1100: attocube). Low temperature PL measurements were performed with a Ti:Sa laser (3900S: Spectra-Physics), a spectrometer (SP-2750: Princeton Instruments), and a CCD camera (PIXIS 400: Princeton Instruments). The wavelength was set at 780 nm, which is short enough to excite electrons and holes above the band gap. The mechanical motion was measured using a Doppler interferometer with a He:Ne laser (MLD_230: Neoark). The Ti:Sa and He:Ne lasers were focused on the resonator by using an objective lens with a numerical aperture of 0.82. Stroboscopic PL measurements were performed with an AOM (TEM-110: Brimrose), which shapes the pump laser to rectangular pulse with a duty ratio of 5 %.

## Acknowledgements

We thank Dr. Y. Matsuzaki, Dr. M. Asano, Dr. S. Houri, Dr. K. Takata, and Dr. S. Adachi for discussions about the theoretical models and fruitful comments on the manuscript. This work is partly supported by MEXT KAKENHI (JP15H05869, JP16H01057)

## Author contributions

R.O., H.O., and H.Y. designed the experiments. R.O. and H.O. fabricated the samples, and developed the experimental setup with support from T.T. and H.G. The measurements and calculations were performed by R.O.. R.O., H.O., T.T. and H.Y. analyzed the data. R.O. and H.Y. wrote the manuscript. All of the authors discussed the results and contributed to the manuscript.

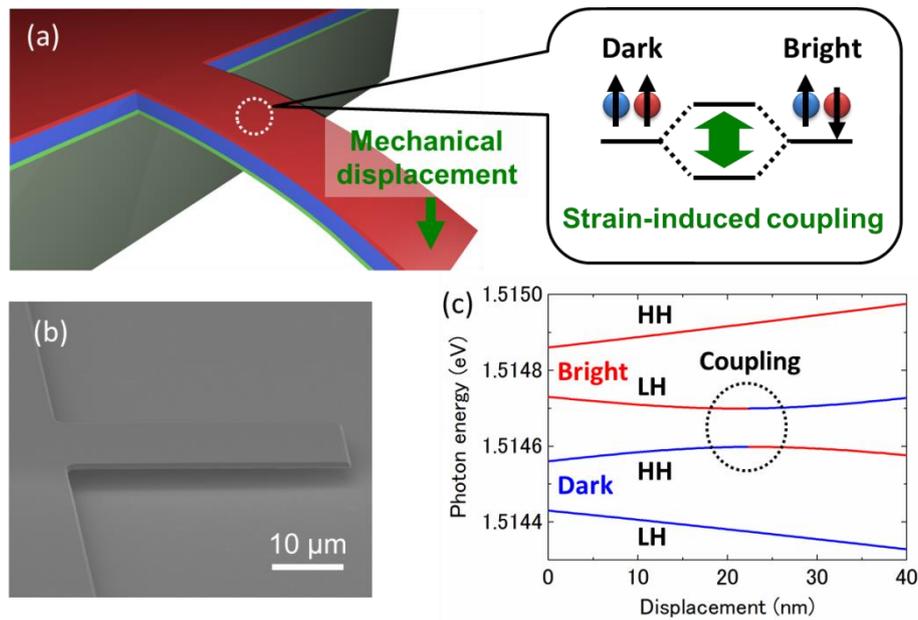

**Figure 1| Schematic and SEM images of the mechanical resonator and energy diagram of bound exciton states. a,** Schematic image of the mechanical resonator and the strain-induced coupling of dark and bright excitons. **b,** SEM image of the fabricated resonator (37 µm long, 20 µm wide, and 600 nm thick). **c,** Energy diagram of dark and bright bound excitons with mechanical displacement calculated by $\boldsymbol{k}\cdot\boldsymbol{p}$ theory. In-plane strain enlarges the splitting of heavy-hole (HH) and light-hole (LH) bands, and generates coupling between dark and bright excitons.

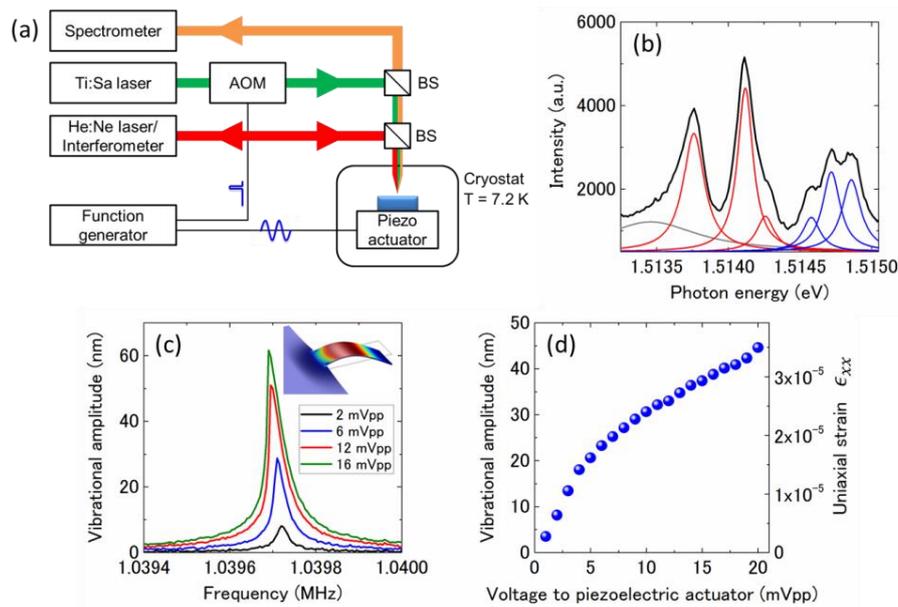

**Figure 2| Optical and mechanical properties of resonator. a,** Experimental setup for characterizing the optical and mechanical properties of a resonator with bound excitons. PL spectra were obtained with a Ti:Sa pump laser, whose oscillating wavelength was 780 nm. Mechanical motion was measured using a Doppler interferometer with a He:Ne laser. The resonator was placed on a piezo actuator, which drove it at the resonance frequency. Experiments were performed in a vacuum at 7.2 K **b,** PL spectrum obtained at the mid-point of the resonator. It is fitted by seven Lorenz functions. Red (blue) peaks are attributed to acceptor (donor) bound excitons. The gray peak is considered to be background. **c,** Frequency response of the 2$^{nd}$ flexural mode of the resonator driven by the piezo actuator. The inset shows the strain distribution calculated by FEM. **d,** Voltage dependence of the vibrational amplitude and the corresponding in-plane strain of the resonator at the resonance frequency.

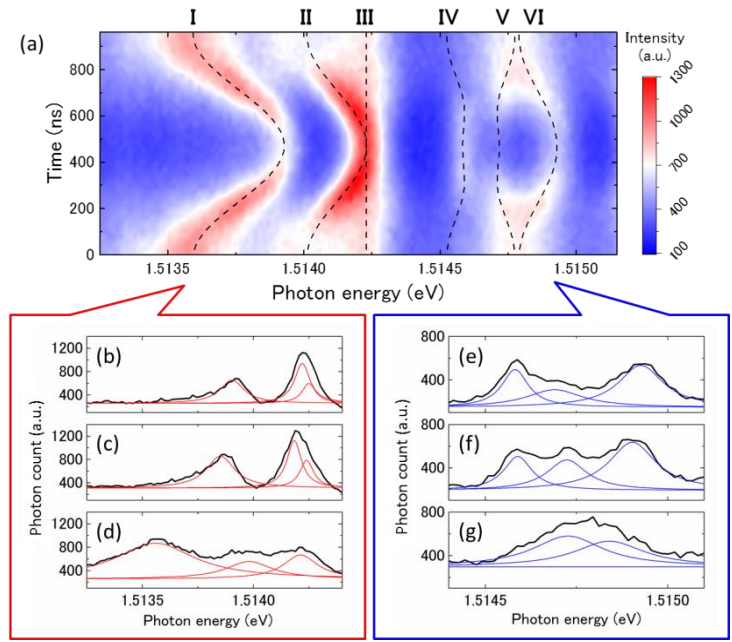

**Figure 3| Stroboscopic PL spectra of acceptor and donor bound excitons. a,** Wide-range stroboscopic PL spectrum. Peaks I to III (IV to VI) are attributed to acceptor (donor) bound excitons. The dashed lines are their fitted peak energies. **b-g,** Cross-sectional PL spectra of acceptor (b-d) and donor (e-g) bound excitons at 0 (d, g), 320 (c, f), and 480 ns (b, e), respectively. Peaks IV and V show the avoided crossing at 320 and 640 ns, respectively, indicating the coherent coupling of the two states. Peak IV is attributed to the dark exciton, which is optically inactive when it detuned from peak V.

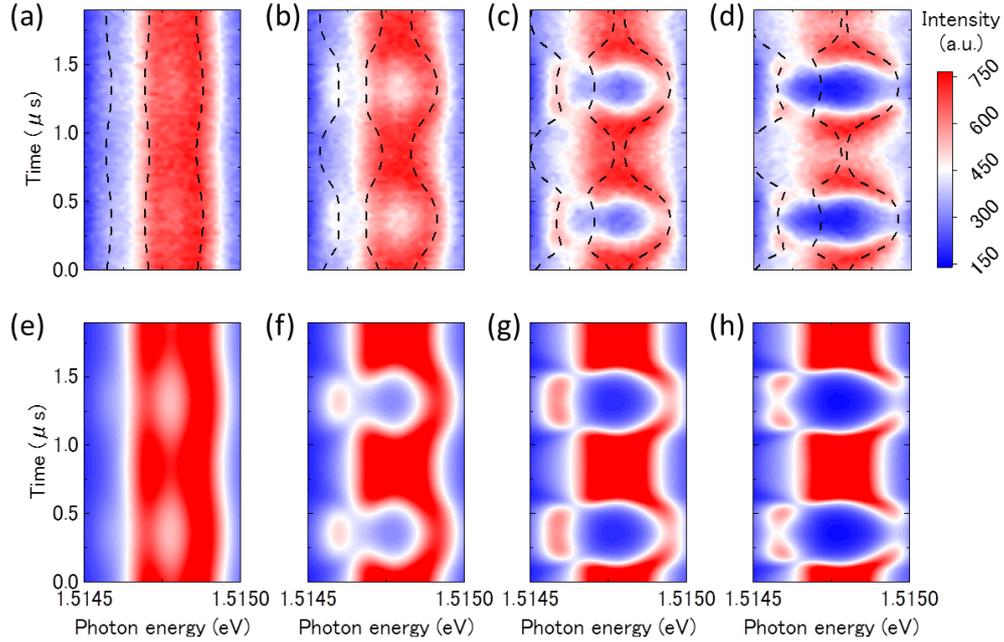

**Figure 4 | Experimental and numerical spectra of dark and bright bound excitons. a-d,** Stroboscopic PL spectra obtained with drive voltages of (**a**) 2, (**b**) 6, (**c**) 12, and (**d**) 16 mV$_{pp}$. The dashed lines are the eigen energies of bound excitons calculated by using the Pikus-Bir Hamiltonian with measured parameters. **e-h,** Numerically calculated PL spectra at the corresponding vibrational amplitudes of **a-d**. The good agreement between the experimental and numerical spectra confirms that the vibrational strain causes the coherent coupling of the dark exciton in the HH band and the bright exciton in the LH band.

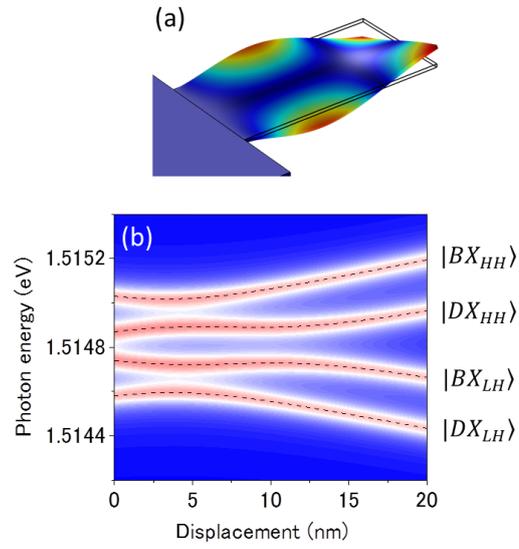

**Figure 5 | Exciton-exciton couplings in four bound states caused by torsional mechanical oscillation. a,** The 2nd order torsional mechanical mode calculated with FEM. **b,** Displacement dependence of the dark and bright bound exciton states calculated using the Pikus-Bir Hamiltonian with the parameters from experiments. $L_\epsilon$ causes the coupling of $|BX_{HH}\rangle$ and $|BX_{LH}\rangle$, and $|DX_{LH}\rangle$ and $|DX_{HH}\rangle$ at 4 nm, while the coupling of $|DX_{HH}\rangle$ and $|BX_{LH}\rangle$ is provided by $M_\epsilon$ at 10 nm

# Supplementary Information

*Strain effects on dark and bright bound excitons*

The wave functions of electrons near the GaAs bandgap are derived from Kane's model, which considers the spin-orbit interaction in $\boldsymbol{k}\cdot\boldsymbol{p}$ theory.

$$H = \frac{k^2}{2m_0} + \frac{\boldsymbol{k}\cdot\boldsymbol{p}}{m_0} + \frac{p^2}{2m_0} + V_{(r)} + \frac{\hbar}{4m_0^2 c^2}(\nabla V_0)\times\boldsymbol{p}\cdot\boldsymbol{\sigma} + \frac{\hbar}{4m_0^2 c^2}(\nabla V_0)\times\boldsymbol{k}\cdot\boldsymbol{\sigma}$$

The first, second, and third terms are standard $\boldsymbol{k}\cdot\boldsymbol{p}$ Hamiltonians, the fourth term is the self-consisted periodic potential $V_{(r)}$, and the fifth and sixth terms are the spin-orbit interaction. $m_0, c,$ and $\hbar,$ are electron mass, light velocity, and Planck's constant, respectively. $\boldsymbol{\sigma}$ is the Pauli spin matrix with components

$$\sigma_x = \begin{bmatrix} 0 & 1 \\ 1 & 0 \end{bmatrix}, \sigma_y = \begin{bmatrix} 0 & -i \\ i & 0 \end{bmatrix}, \sigma_z = \begin{bmatrix} 1 & 0 \\ 0 & -1 \end{bmatrix}$$

This Hamiltonian provides eight electron states with the following wave functions.

$$|u_E^U\rangle = |S\uparrow\rangle$$
$$|u_E^D\rangle = |S\downarrow\rangle$$
$$|u_{HH}^U\rangle = \frac{-1}{\sqrt{2}}|(X+iY)\uparrow\rangle$$
$$|u_{HH}^D\rangle = \frac{1}{\sqrt{2}}|(X-iY)\downarrow\rangle$$
$$|u_{LH}^U\rangle = \frac{-1}{\sqrt{6}}|(X+iY)\downarrow\rangle + \sqrt{\frac{2}{3}}|Z\uparrow\rangle$$
$$|u_{LH}^D\rangle = \frac{1}{\sqrt{6}}|(X-iY)\uparrow\rangle + \sqrt{\frac{2}{3}}|Z\downarrow\rangle$$
$$|u_{SO}^U\rangle = \frac{1}{\sqrt{3}}|(X-iY)\uparrow\rangle - \frac{1}{\sqrt{3}}|Z\downarrow\rangle$$
$$|u_{SO}^D\rangle = \frac{1}{\sqrt{3}}|(X+iY)\downarrow\rangle + \frac{1}{\sqrt{3}}|Z\uparrow\rangle$$

$|u_E\rangle, |u_{HH}\rangle, |u_{LH}\rangle, |u_{SO}\rangle$ are the wave functions of electrons in conduction, heavy hole (HH), light hole (LH), and spin-orbits split-off bands, where $|S\rangle$ is an *s*-shaped function and $|X\rangle, |Y\rangle,$ and $|Z\rangle$ are *p*-shaped functions, and

↑ (↓) indicates the spin up (down) state. Electron and hole pairs of $|u_E^U\rangle$ and $|u_{HH}^D\rangle$ ($|u_E^D\rangle$ and $|u_{HH}^U\rangle$) form bright excitons ($|BX_{HH}\rangle$), and those of $|u_E^U\rangle$ and $|u_{HH}^U\rangle$ ($|u_E^D\rangle$ and $|u_{HH}^D\rangle$) form dark excitons ($|DX_{HH}\rangle$). The energies of the dark excitons are lower than those of the bright excitons due to the exchange energy of an electron and a hole ($\varepsilon_{ex}$). We assume that the potential gradients of donors are much broader than the lattice constant so that the effective mass approximation well describes the bound exciton energies[S1].

The strain effect of each state is characterized by the Luttinger-Kohn and Pikus-Bir Hamiltonian for bases of ($|u_E^U\rangle$, $|u_E^D\rangle$, $|u_{HH}^U\rangle$, $|u_{LH}^U\rangle$, $|u_{LH}^D\rangle$, $|u_{HH}^D\rangle$, $|u_{SO}^U\rangle$, $|u_{SO}^D\rangle$))

$$\begin{pmatrix}
E_g+r & 0 & 0 & 0 & 0 & 0 & 0 & 0 \\
0 & E_g+r & 0 & 0 & 0 & 0 & 0 & 0 \\
0 & 0 & p+q & l & m & 0 & -\frac{1}{\sqrt{2}}l & -\sqrt{2}m \\
0 & 0 & l^* & p-q & 0 & m & \sqrt{2}q & \sqrt{\frac{3}{2}}l \\
0 & 0 & m^* & 0 & p-q & -l & \sqrt{\frac{3}{2}}l^* & -\sqrt{2}q \\
0 & 0 & 0 & m^* & -l^* & p+q & \sqrt{2}m^* & -\frac{1}{\sqrt{2}}l^* \\
0 & 0 & -\frac{1}{\sqrt{2}}l^* & \sqrt{2}q & \sqrt{\frac{3}{2}}l & \sqrt{2}m & p-\Delta & 0 \\
0 & 0 & -\sqrt{2}m^* & \sqrt{\frac{3}{2}}l^* & -\sqrt{2}q & -\frac{1}{\sqrt{2}}l^* & 0 & p-\Delta
\end{pmatrix}$$

$$r = r_k + r_\epsilon$$
$$p = p_k + p_\epsilon$$
$$q = q_k + q_\epsilon$$
$$l = l_k + l_\epsilon$$
$$m = m_k + m_\epsilon$$
$$r_k = \varepsilon_c + \frac{\hbar^2}{2m_e^*}(k_x^2 + k_y^2 + k_z^2)$$
$$p_k = \varepsilon_v - \frac{\hbar^2}{2m_0}\gamma_1(k_x^2 + k_y^2 + k_z^2)$$

$$q_k = -\frac{\hbar^2}{2m_0}\gamma_2(k_x^2 + k_y^2 - 2k_z^2)$$

$$l_k = \frac{\hbar^2}{2m_0}\gamma_3 2\sqrt{3}(k_x - ik_y)k_z$$

$$m_k = \frac{\hbar^2}{2m_0}\sqrt{3}[\gamma_2(k_x^2 - k_y^2) - i2\gamma_3 k_x k_y]$$

$$r_\epsilon = a_c(\epsilon_{xx} + \epsilon_{yy} + \epsilon_{zz})$$

$$p_\epsilon = a_v(\epsilon_{xx} + \epsilon_{yy} + \epsilon_{zz})$$

$$q_\epsilon = \frac{b}{2}(\epsilon_{xx} + \epsilon_{yy} - 2\epsilon_{zz})$$

$$l_\epsilon = -d(\epsilon_{xz} - i\epsilon_{yz})$$

$$m_\epsilon = -\frac{\sqrt{3}}{2}b(\epsilon_{xx} - \epsilon_{yy})$$

$\gamma_{1,2,3}$ are Luttinger parameters and $a_c, a_v, b,$ and $d$ are hydrostatic and shear deformation potentials. $\epsilon_{ij}$ is the strain tensor component. $E_g$ is the bandgap energy of GaAs. The cross terms between conduction and valence bands are negligible around Γ-point ($k_x, k_y, k_z = 0$), and so the strain effects for the bound exciton states ($|BX_{HH}\rangle, |BX_{LH}\rangle, |DX_{LH}\rangle, |DX_{HH}\rangle$) are described as follows.

$$\begin{pmatrix} E_g^* + P_\epsilon + Q_\epsilon & L_\epsilon & M_\epsilon & 0 \\ L_\epsilon^* & E_g^* + P_\epsilon - Q_\epsilon & 0 & M_\epsilon \\ M_\epsilon^* & 0 & E_g^* + P_\epsilon - Q_\epsilon - \varepsilon_{ex} & -L_\epsilon \\ 0 & M_\epsilon^* & -L^* & E_g^* + P_\epsilon + Q_\epsilon - \varepsilon_{ex} \end{pmatrix}$$

$$P_\epsilon = (a_c - a_v)(\epsilon_{xx} + \epsilon_{yy} + \epsilon_{zz})$$

$$Q_\epsilon = -b(\epsilon_{xx} + \epsilon_{yy} - 2\epsilon_{zz})/2$$

$$L_\epsilon = d(\epsilon_{xz} - i\epsilon_{yz})$$

$$M_\epsilon = \sqrt{3}b(\epsilon_{xx} - \epsilon_{yy})/2$$

where $E_g^*$ takes account of the binding energies. $\varepsilon_{ex}$ is the exchange energy of an electron and a hole. The diagonal terms of $Q_\epsilon$ cause the energy splitting of HH and LH bands. The off-diagonal terms of $M_\epsilon$ cause the couplings of $|BX_{HH}\rangle$ and $|DX_{LH}\rangle$ ($|BX_{LH}\rangle$ and $|DX_{HH}\rangle$).

In the experiments, there are two kinds of strains, namely the static strain from the lattice mismatch of GaAs and $Al_{0.3}Ga_{0.7}As$, and the dynamic strain caused by the vibration. The former is an isotropic in-plane strain ($\epsilon_{xx} \cong \epsilon_{yy}$), which causes the intrinsic energy difference between the LH and HH bands ($\varepsilon_{LH} = 2 \times Q_\epsilon$). However, the mixing terms from the intrinsic strain are negligible because $\epsilon_{xx}$ and $\epsilon_{yy}$ cancel each other out ($Q_\epsilon \gg M_\epsilon$). On the other hand, the mechanical motion in the cantilever structure as shown in Fig. 1(a) generates a uniaxial strain ($\epsilon_{xx} \gg \epsilon_{yy}$), which provides both the energy splitting of the LH and HH bands and the mixing of the dark and bright excitons. We calculated the strain tensors of the 2nd flexural mode numerically using FEM at the measured position of the mechanical resonator. The parameters $\varepsilon_{ex}, b,$ and $d$ were $2.9 \times 10^{-4}$, -2.2, -5.4 eV based on references[18,28], and $E_g^*$, $\varepsilon_{LH}$, and $a$ were extracted as 1.51486, $1.7 \times 10^{-4}$, and -0.5 eV by fitting the experiments. The numerical solutions of the dark and bright exciton energies obtained with $\boldsymbol{k} \cdot \boldsymbol{p}$ theory are shown in Fig. 1(d) and Fig. 4(a)-(d).

*Numerical calculation of PL spectra of bound excitons*

PL spectra are calculated from the population of bright excitons, which are described by the creation-annihilation operators in harmonic oscillator approximation. In this case, the creation-annihilation operator of each state is derived from the steady states of the Heisenberg equation of motion with the Pikus-Bir Hamiltonian.

$$\begin{pmatrix} \omega_{BH} - \omega & L_\epsilon & M_\epsilon & 0 \\ L_\epsilon^* & \omega_{BL} - \omega & 0 & M_\epsilon \\ M_\epsilon^* & 0 & \omega_{DL} - \omega & -L_\epsilon \\ 0 & M_\epsilon^* & -L_\epsilon^* & \omega_{DH} - \omega \end{pmatrix} \begin{pmatrix} b_{HH} \\ b_{LH} \\ d_{LH} \\ d_{HH} \end{pmatrix} = \begin{pmatrix} \lambda_{BH} \\ \lambda_{BL} \\ \lambda_{DL} \\ \lambda_{DH} \end{pmatrix}$$

$$\omega_{BH} = E_g^* + P_\epsilon + Q_\epsilon - i\gamma_{BH}$$
$$\omega_{BL} = E_g^* + P_\epsilon - Q_\epsilon - i\gamma_{BL}$$
$$\omega_{DL} = E_g^* + P_\epsilon - Q_\epsilon - \varepsilon_{ex} - i\gamma_{DL}$$
$$\omega_{DH} = E_g^* + P_\epsilon + Q_\epsilon - \varepsilon_{ex} - i\gamma_{DH}$$

Here, $\gamma$, $b(d)$, and $\lambda$ are the linewidth, annihilation operator, and excitation of each state, respectively. We assume that the excitation ratios are the same in the four states. The PL intensities are derived from $b_{HH}^\dagger b_{HH} + 0.88 \times b_{LH}^\dagger b_{LH}$, where we consider the collection efficiency of an objective lens with a numerical aperture of 0.82. Two-thirds of the photons are emitted in the in-plane direction from LH band-excitons.

Mechanical bending additionally modulates the PL intensity via the piezo potential effect[S1], which is proportional to the magnitude of the in-plane strain. The piezo potential effect causes a potential gradient and modifies carrier distributions, because the magnitude of the in-plane strain varies linearly in the thickness direction. It reduces the PL intensity of all bound excitons, i.e. acceptor and donor bound excitons, at a large mechanical displacement. We take account of the time-varying intensity reduction from the experimentally obtained ensemble photon modulation from all bound excitons.

*Position dependence of PL spectra*
Figure S1(b) shows the position dependence of PL spectra obtained in the length direction of the cantilever, i.e. along the dotted line in Fig. S1(a). Around the clamping point of the resonator, the exciton peaks are broadened because of the large non-uniform strain imposed by the lattice-mismatched sacrificial layer. Such non-uniform strain is released around the middle of the resonator, and so sharp peaks originating from the bound exciton states are observed. On the basis of this position dependence of the PL spectra, we adopted the 2nd order flexural mode to efficiently induce in-plane strain for bound excitons, whereas the fundamental mode generates the strain at the bottom of the resonator where each exciton state is not resolved in the PL spectra. The PL spectra in Figs. 3 and 4 were obtained at a position 22 μm from the bottom.

*Tailoring coupling strength and energy shifts via strain engineering*
The $\mathbf{k}\cdot\mathbf{p}$ model calculation indicates that the coupling strength and energy shifts can be independently manipulated by the strain distribution, i.e. the magnitude and anisotropy of the strain tensors. To enhance $g_{DB}$, we need to maximize $M_\epsilon$ and inject a large vibrational amplitude. However, a large vibrational amplitude causes the energy detuning of dark and bright excitons, and so we also need to minimize $P_\epsilon$ and $Q_\epsilon$ to involve the strong interaction in two states. A large $M_\epsilon$ and a small $P_\epsilon$ and $Q_\epsilon$ are generated by the anisotropic strain,
$$\epsilon_{xx} = -\epsilon_{yy}$$
$$\epsilon_{zz} = 0,$$
which result in the following diagonal and off-diagonal terms.
$$M_\epsilon = \sqrt{3}b\epsilon_{xx}$$
$$P_\epsilon = Q_\epsilon = 0$$
Such anisotropic strain can be realized in a drum-shaped resonator as shown in Fig. S2(a), whereas a cantilever dominantly causes the uniaxial strain discussed in the main text. Here, we adopted two orthogonal modes, mode H and mode V as shown in Fig. S2(b)-(c), and chose their cross point where $\epsilon_{xx} = -\epsilon_{yy}$ and $\epsilon_{zz} = 0$. Figure S2(d)-(i) shows $\epsilon_{xx}$, $\epsilon_{yy}$, and $\epsilon_{zz}$ in mode H and mode V calculated by FEM. The matrix elements (μeV/nm) are derived from the sum of two strain tensors and the parameters in the experiments.

$$P_\epsilon = -0.22$$
$$Q_\epsilon = -0.75$$
$$L_\epsilon = 2.31 - 1.27i$$
$$M_\epsilon = -45.3$$

Although numerical calculation does not perfectly cancel $P_\epsilon$ and $Q_\epsilon$, $M_\epsilon$ is two orders of magnitude larger than they are. We calculated the PL spectrum from dark and bright excitons with the experimentally available vibrational amplitude as shown in Fig. S2(j). $g_{DB}$ reaches 1 meV, which is much larger than the exciton linewidths ($\gamma_B, \gamma_D$), the initial energy difference of dark and bright excitons ($\varepsilon_{ex} - \varepsilon_{LH}$), and the strain-induced energy detuning ($2Q_\epsilon$). The cooperativity ($C = 4g_{DB}^2/\gamma_B\gamma_D$) is about 800, and thus allows the coherent manipulation of dark and bright bound exciton states at the single quantum level.

*Exciton-exciton coupling in four bound exciton states*

The coherent couplings of the HH and LH bands emerge when the off-diagonal terms of $L_\epsilon$ become large, which is due to the shear strain of $\epsilon_{xz}$ and $\epsilon_{yz}$. Torsional mechanical vibration, as shown in Fig. 5(a), generates a large shear strain around the side edges of the resonator. We calculated the matrix elements at a position 100 nm from the side edge.

$$P_\epsilon = 0.46$$
$$Q_\epsilon = 7.35$$
$$L_\epsilon = 9.99 + 0.18i$$
$$M_\epsilon = -6.77$$

Here, $L_\epsilon$ is large enough to cause exciton-exciton coupling in the HH and LH bands. Instead of the PL intensity, we calculate the population of four bound exciton states ($b_{HH}^\dagger b_{HH} + b_{LH}^\dagger b_{LH} + d_{HH}^\dagger d_{HH} + d_{LH}^\dagger d_{LH}$) with strain-induced interactions. We assume that the excitation efficiencies for each state are the same, and that the correction efficiencies and piezo potential effects can be neglected. Here, $\varepsilon_{LH}$ is -1.6 × 10⁻⁴ eV. Avoided crossings between dark-dark and bright-bright excitons are observed together with those between dark-bright excitons.

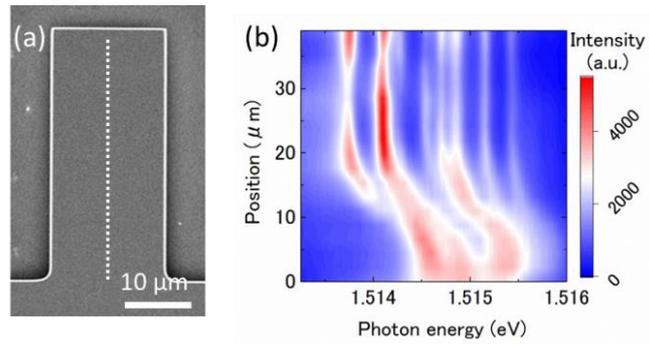

**Figure S1| Position dependence of PL spectra. a,** Top view of the SEM image of the resonator. **b,** PL spectra obtained along the dotted line in **a**. PL spectra are broadened and interfused around the bottom of the resonator due to the non-uniform strain imposed by the sacrificial layer. The non-uniform strain is released around the middle, therefore the PL peaks of the acceptor and donor bound excitons can be distinguished.

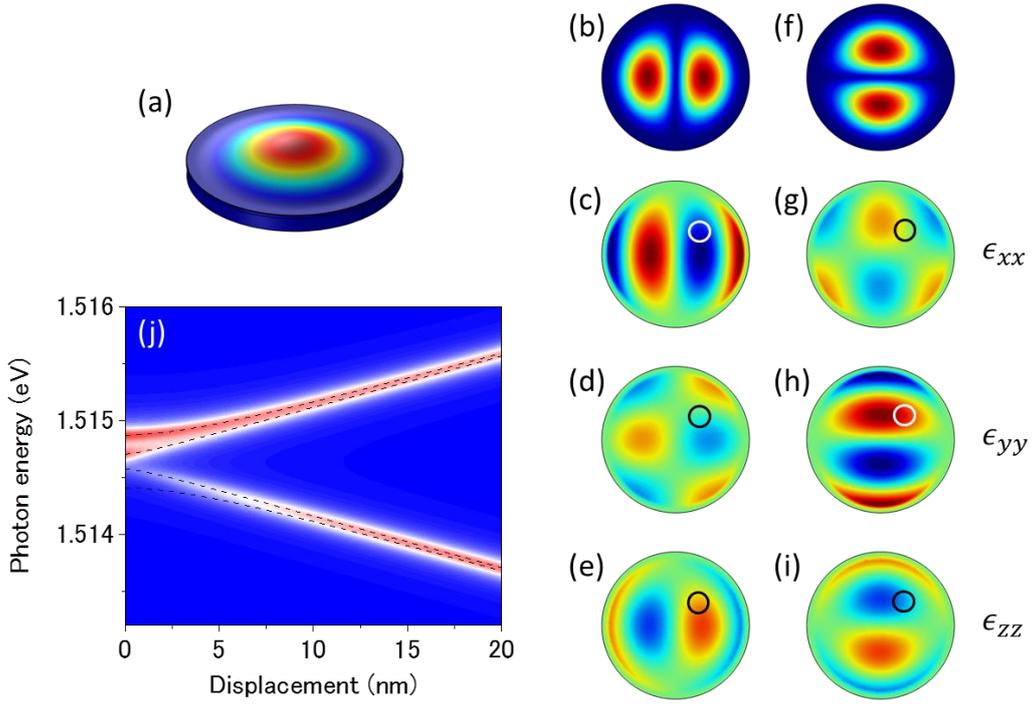

**Figure S2| Strain tensors and dark-bright coupling in drum-shaped resonator. a,** Schematic image of a drum-shaped resonator. The radius and thickness are 20 µm and 600 nm, respectively. **b-i,** Spatial distribution of amplitude (**b,f**), $\epsilon_{xx}$ (**c,g**), $\epsilon_{yy}$ (**d,h**), and $\epsilon_{zz}$ (**e,i**) in mode H (**b-e**) and mode V (**f-i**). **j,** Calculated PL spectra from dark and bright bound excitons with the strain tensors at the circled position in **b-i**. It clearly shows the strong coupling with a $g_{DB}$ of 1 meV, where the detuning of the dark and bright excitons is not significantly changed by mechanical displacement.